\documentclass[12pt]{iopart}
\usepackage{graphicx}
\usepackage{caption}
\begin{document}
\title{Studying Intermediate $p_T$ Hadron Production with Fluctuations}

\author{Thomas Peitzmann}

\address{Utrecht University, P.O. Box 80000, 3508-TA Utrecht, The Netherlands}

\ead{t.peitzmann@phys.uu.nl}

\begin{abstract}
Mechanisms for particle production at intermediate $p_T$ in nuclear collisions at RHIC are discussed, emphasizing the differences in associated jet-like correlations between color-neutral and colored production. An alternative production mechanism involving both recombination and fragmentation is suggested, which might simultaneously lead to an enhancement of baryons and to jet-like correlations. To gain more insight into the relative importance of different mechanisms a study of constrained distributions of associated multiplicity is proposed. In a simple model it is shown that these multiplicity distributions may change significantly, if the nature of the production mechanism fluctuates from event to event.
\end{abstract}.


% Uncomment for Submitted to journal title message
%\submitted

% Comment out if separate title page not required
% \maketitle

\section{Introduction}

Recent results from the Relativistic Heavy Ion Collider on the suppression of high transverse momentum hadrons in central heavy ion collisions at $\sqrt{s_{_{NN}}} = 200\, \mathrm{GeV}$~\cite{phenix:quench}, which are being attributed to energy loss of hard scattered partons in the dense medium created in these collisions \cite{gyulassy:quench,baier:quench}, have received a lot of attention. One has also clearly seen two-particle correlations at high $p_T$ which provide a direct hint of jet structures, i.e. of the hard scattering nature of particle production \cite{star:quench}. Relative to a high $p_T$ trigger particle a small-angle correlation of associated particles is observed, which can be attributed to the fragmentation of a hard scattered parton. In p+p collisions and peripheral Au+Au collisions one also observes a back-to-back correlation, which is seen as the signature of the balancing jet. This back-to-back jet disappears in central Au+Au collisions, again in line with jet quenching.

In the same reactions one has also observed striking features of collectivity, most importantly a strong elliptic anisotropy persisting out to very high transverse momenta~\cite{star:flow}. The strength of this anisotropy $v_2$ has a peculiar dependence on the particle species~\cite{star:flowlam}. Also the suppression of inclusive hadrons appears to be distinctively different for different hadrons, with little or no suppression for baryons in an intermediate $p_T$ region~\cite{phenix:bar,star:flowlam}. It has been proposed \cite{fries:reco,greco:reco} that these species dependences in the intermediate $p_T$ regime ($2 - 6 \, \mathrm{GeV}/c$) may be explained by a collective production mechanism, namely \textit{recombination} or \textit{coalescence} (called ``ReCo'' below).

In the ReCo mechanism hadrons are made through assembly of existing quarks or antiquarks. If the phase space density of (anti-)quarks is sufficiently high, this provides for a natural way of enhancing baryons relative to mesons at a given momentum, as one adds up three quark momenta instead of two, and also helps making this production mechanism more favorable relative to jet fragmentation. ReCo is further enhanced compared to fragmentation due to the penalty the latter receives because a colored object (the parton) tries to pull away from another color source and loses energy in the field. ReCo deals with the formation of color singlets which do not suffer such interaction effects.

In this basic version of ReCo it is clear that the produced hadrons leaving the interaction zone as singlets will not have any associated particles as in jet fragmentation. It has been argued \cite{fries:qm2004} that, as there are correlations between partons in a quark-gluon-plasma, there should also be remnants of these between the emerging hadrons. While this is certainly true in principle, it is very unlikely that this will produce similar correlation patterns as for jets in the vacuum. The strongest correlations in the QGP should be remnants of bound states which are dominantly at low relative momentum and small distance. Most of the correlated partons might therefore end up in one hadron as such near-by partons are just required for ReCo, and only a small fraction of them might contribute to correlations between different hadrons. While this issue is not even qualitatively under control -- much less quantitatively, I consider it very unlikely that such correlations would yield jet-like structures. So, a production mechanism like ReCo, which draws part of its importance from the fact that there is no fragmentation-like ``color penalty'', will lead to a modified and most likely strongly reduced jet-like correlation. In fact, this should be true for any similar mechanism: while it is e.g. possible to account for pion, kaon and proton spectra simultaneously with a suitably adapted hydrodynamic parameterization \cite{tp:hydro}, this would also have no natural way of explaining correlation structures. It should be noted that this points to a consistency problem, when we simultaneously invoke hadron hydrodynamics to explain low $p_T$ spectra and a different mechanism to account for intermediate $p_T$ behavior. Just extrapolating e.g. the hydrodynamic description of the proton spectra from low $p_T$ to intermediate $p_T$ leaves little room for another production mechanism. It is an open question how the transition between the two regimes can be made.

The ReCo processes invoked for explaining the observations in heavy ion collisions rely on directly producing color singlets. It may be important to also consider other processes relaxing this constraint. In particular, for the production of baryons one might consider the formation of a diquark by a ReCo mechanism, which is then emitted and fragments yielding a leading baryon. The diquark receives a similar enhancement as a meson from the summing of quark momenta. It will still fragment and thereby the baryon will not carry all the momentum, there is, however, an enhancement by the color factor 9. So depending on temperature, phase space density and the fragmentation function of the diquark this process may actually be important for baryon production. The ratio of baryons produced by such a mechanism to mesons from ordinary ReCo assuming an exponential spectrum may be approximately estimated as:

%\begin{equation}
%\frac{N_{baryon}}{N_{meson}} = 9 \cdot \exp\left[ -\frac{p_T}{T} \left( \frac{1}{\langle z \rangle} - 1 \right) \right].
%\label{eq1}
%\end{equation}
For $p_T = 3 \, \mathrm{GeV}/c$ and an effective temperature $T = 350 \, \mathrm{MeV}$ this reaches the order of 1, if the leading baryon on average carries a momentum fraction of $\langle z \rangle > 0.8$. This simple estimate suggests that such a process should yield a non-negligible contribution to baryon production at intermediate $p_T$. The interesting feature of this process is that the leading baryon would have associated particles from the fragmentation and thus contribute to a jet signature.

I will not follow this up in the present paper. It should, however, be evident that it is of importance to understand the underlying production processes. I will in the following distinguish between \textit{color-neutral production} and \textit{colored production}, where the former does not lead to jet structures while the latter does. The original ReCo mechanism is the most prominent color-neutral example and fragmentation belongs to the colored category. If both classes of processes occur in heavy ion reactions, in some events a particle at intermediate $p_T$ would have associated particles, in others it wouldn't. This should in general lead to very different particle number fluctuations close to a intermediate $p_T$ particle. The purpose of this paper is to study such fluctuations with the goal to find observables which can discriminate between the different mechanisms.

\section{The Model}

The basic idea followed in this paper is to study multiplicity distributions (MD) in limited regions of phase space where a jet may be located. It was seen in two-particle correlation analysis in p+p and Au+Au collisions at 200 GeV that within a narrow angular cone around a high $p_T$ trigger particle one can find correlated associated particles~\cite{star:quench}. I will therefore study the multiplicity associated with a high $p_T$ particle. For this purpose I will use two windows in azimuth, namely $| \Delta \phi | < 30 ^{\circ}$ and $| \Delta \phi | < 90 ^{\circ}$. In real data analysis one would also constrain the analysis within windows of $\Delta \eta$ as used e.g. by the STAR experiment in the above publication. To have a significant jet signal one would use appropriate $p_T$ thresholds for the particles, e.g. $p_T > 4 \, {\mathrm GeV}/c$ for the trigger particle and $p_T > 2 \, {\mathrm GeV}/c$ for the associated particles. This would imply that one has to deal with relatively small numbers of particles per event. (E.g. in p+p collisions simulated by HIJING \cite{hijing} the corresponding associated multiplicity is approximately 1.) I will call the multiplicity in the smaller interval $I$ and the multiplicity in the larger one $J$.

I will now assume that all associated particles from a original jet selected by the leading trigger particle are observed in the smaller angular window ($| \Delta \phi | < 30 ^{\circ}$), which is approximately in line with observations in p+p collisions  \cite{star:quench}. The multiplicity distribution can be obtained e.g. from HIJING - it has been parameterized with a Gaussian distribution:
\begin{equation}
\label{eq2}
P^{pp}(I) = \frac{1}{\sqrt{2 \pi} \sigma^{pp}} \cdot \exp\left[ - \frac{\left( I - \mu^{pp} \right)^2} {2 (\sigma^{pp})^2} \right]
\end{equation}
resulting in a mean $\mu^{pp} = 1.18$ and width $\sigma^{pp} = 0.55$.

For A+A collisions I will first assume that a trigger particle should be associated with particles according to the same multiplicity distribution as in p+p. In addition there is a background which is in the most simple approximation random relative to the trigger particle. It is well known that elliptic flow would lead to a correlation in this background distribution, but I will ignore this for the moment. Even without correlations this background will show number fluctuations, and the higher the background, the more difficult it will be to get information on the fluctuations of the associated multiplicity. Furthermore, there will always be fluctuations in the geometry of the collision resulting in additional multiplicity fluctuations.

\begin{figure}[htb]
\begin{minipage}{18pc}
\includegraphics{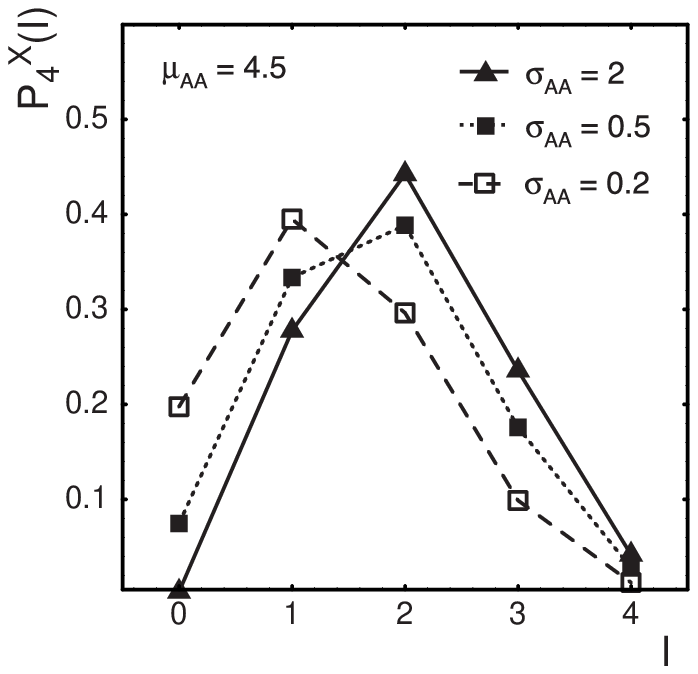} 
\caption{Constrained multiplicity distributions for different background multiplicity distributions. (See text for discussion. \label{fig1}}
\end{minipage}\hspace{2pc}%
\begin{minipage}{18pc}
\includegraphics{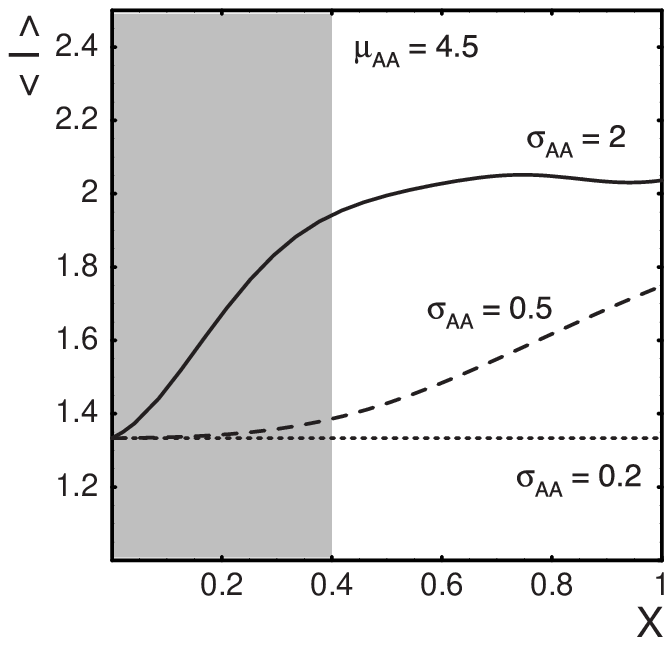} 
\caption{\label{fig2}Mean multiplicity $<I>$ for different background multiplicity distributions as in Fig.~\ref{fig1}.}
\end{minipage} 
\end{figure}

To avoid some of these complications I will study \textit{constrained multiplicity} distributions: Keeping the multiplicity $J$ ($| \Delta \phi | < 90 ^{\circ}$) fixed, I will look at the distribution $P_J(I)$ ($| \Delta \phi | < 30 ^{\circ}$). This distribution contains essentially information on how a given number of particles is distributed in angular space. Our assumption on the MD in p+p would mean that
\begin{equation}
\label{eq3}
P^{pp}_J(I) = \left\{ \begin{array}{l@{\quad:\quad}l} 1 & I = J  \\
0 & I \neq J. \end{array} \right.
\end{equation}
In the other extreme, for random distributions, this multiplicity would be described by a Binomial distribution with the single probability $p = 1/3$:
\begin{equation}
\label{eq4}
P^{ran}_J(I) = \left( \begin{array}{c} I \\
J \end{array} \right) \cdot p^{I} (1-p)^{J-I}.
\end{equation}
In a real A+A event the MD will contain both the components from the jet distribution known from p+p and a background component, which I will assume to be random. The relative importance
of the two components will depend on the background multiplicity and thus on the centrality of the collision. To obtain the MD in this case one would have to fold the p+p distribution and the random background distribution with the background MD $P^{AA}(J)$. This yields:
\begin{eqnarray}
\label{eq5}
P^{AA}_J(I) = \sum_{K=0}^{\min{I,J}} P^{pp}(K) \cdot P^{ran}_{J-K}(I-K) \cdot P^{AA}(J-K) \nonumber \\ \, / \sum_{L=0}^{J} \sum_{K=0}^{\min{L,J}} P^{pp}(K) \cdot P^{ran}_{J-K}(L-K) \cdot P^{AA}(J-K), \, \,
\end{eqnarray}
where the sum in the denominator is purely for normalization. The shape of this distribution essentially depends on the background MD. The more events with high background multiplicities dominate, the closer the distribution gets to the random case. This has been investigated using a particular arbitrary shape of the background MD, a Gaussian with mean $\mu_{AA}$ and sigma $\sigma_{AA}$. The effects of this are illustrated in Fig.~\ref{fig1} for the special case $J \equiv 4$, where $\mu_{AA} = 4.5$ is used for different values of $\sigma_{AA}$. For a small width ($\sigma_{AA} = 0.2$) the random distribution dominates, and a Binomial distribution is recovered (open squares). When increasing the width, lower background multiplicity gets more and more important, and the jet contribution is obviously stronger, which leads to an increase of the mean of the distribution. Obviously, if a jet correlation peak is observed in the associated particles, the mean must be significantly different from the one of a Binomial. Lowering the mean $\mu_{AA}$ of the background MD will naturally also lead to to a larger mean of the distribution $P^{AA}_J(I)$.
The model can be extended by assuming that only a fraction $X$ of the events shows a jet correlation in associated particles and that the remaining fraction has only random background. In this case the full distribution would be described by:
\begin{eqnarray}
\label{eq6}
P^{AA}_J(I) = (1-X) \cdot P^{ran}_J(I)  \nonumber \\ 
\, \, + X \cdot \sum_{K=0}^{\min{I,J}} P^{pp\prime}(K) \cdot P^{ran}_{J-K}(I-K) \cdot P^{AA}(J-K)  \nonumber \\
 \, \, / \sum_{L=0}^{J} \sum_{K=0}^{\min{L,J}} P^{pp\prime}(K) \cdot P^{ran}_{J-K}(L-K) \cdot P^{AA}(J-K),
\end{eqnarray}
where $P^{pp\prime}(I)$ is a modified jet MD with 
$\mu^{pp\prime} = \mu^{pp}/X$ and $\sigma^{pp\prime} = \sigma^{pp}/X$. This scaling of the distribution is a na{\"{\i}}ve way of implementing a stronger jet peak. Fig.~\ref{fig2} illustrates how variations in the parameters change the mean of the distributions. For the same $\mu_{AA}$ as above and a large $\sigma_{AA}$ one can obtain a relatively constant mean of the final distribution as a function of $X$ for $X > 0.4$. For smaller values of $X$ this must necessarily break down, because in the limit $X = 0$ one must recover a Binomial which has to yield $<I> = 4/3$.

\begin{figure}[htb]
\begin{minipage}{18pc}
\includegraphics{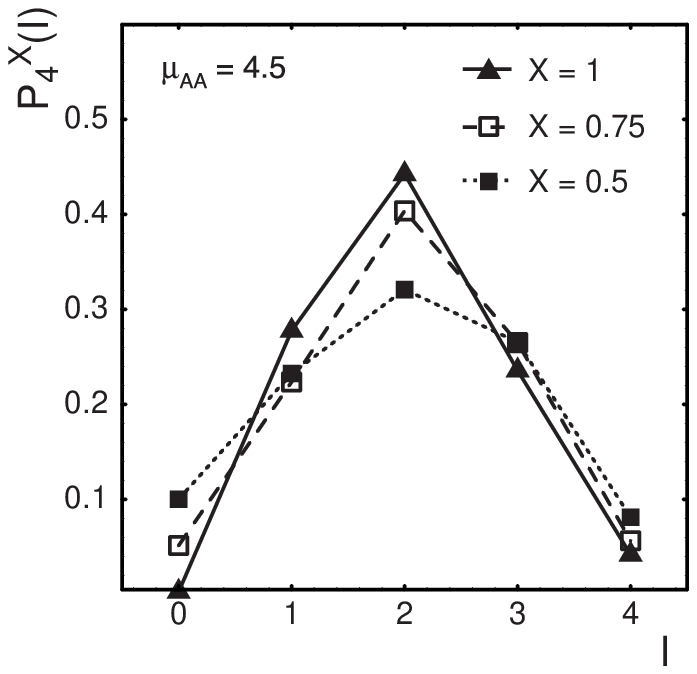} 
\parbox{18pc}{\caption{Constrained multiplicity distributions for different fractional contribution $X$ of jet events. \label{fig3}}}%}
\end{minipage} \hspace{2pc}%
\begin{minipage}{18pc}
\includegraphics{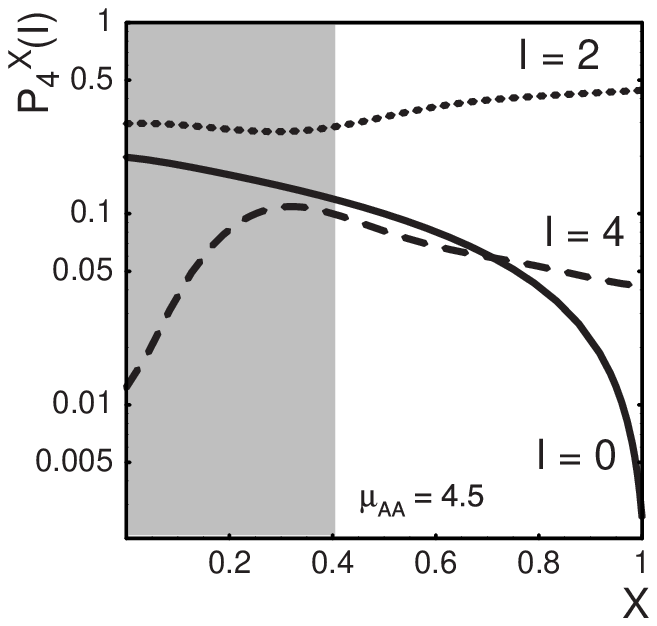} 
\caption{\label{fig4} Probability for different values of the multiplicity $I$ as a function of the fractional contribution $X$ of jet events.}
\end{minipage} 
\end{figure}
Many other parameter settings yield a dominance of the random background and would thus correspond to no visible jet-like correlation. For comparison to real data one would have to tune the parameters more to reach a reasonable agreement, but in this context I will use one particular parameter set for illustration. I will study the properties of these distributions further for the interesting example of $\mu_{AA} = 4.5$ and $\sigma_{AA} = 2$. The mean of the distributions, as shown above, are independent of $X$ in a certain range and could thus be responsible for the same average jet structure, but it is of interest to study how the distributions themselves behave for different $X$. This is illustrated in Figs.~\ref{fig3} and \ref{fig4}. Fig.~\ref{fig3} shows examples of constrained MD for different $X$. One can nicely see that the mean of the distributions is essentially unchanged, but that they get significantly wider. Quantitatively this can be seen better in Fig.~\ref{fig4} which shows the probabilities of certain multiplicities $I$ as a function of $X$. In the interesting range ($X > 0.4$) one can see that especially $P(I=0)$ shows a dramatic relative variation, if $X$ is decreased from 1. This can be easily understood qualitatively. If there is an associated jet in every event, it is very unlikely to observe no additional particle around the trigger particle. This probability increases strongly, if some events have just a random particle distribution. The model employed here ignores completely possible correlations in the background multiplicity. Sources of such correlations might e.g. be resonance decays or elliptic flow. While resonances would definitely be important at low $p_T$, it is assumed that their influence is much less vital at the momenta relevant here. Elliptic flow, and in particular fluctuations in the flow strength may play a significant role. These effects, and possible other influences remain to be studied.

\section{Summary}
The different signatures of colored and color-neutral particle production with respect to jet-like correlations have been discussed. While color-neutral processes like those underlying recombination/coalescence models can explain an enhancement of baryons relative to mesons at intermediate $p_T$, they can so far not easily account for the jet-like structures observed when triggering on high $p_T$ particles. A \textit{hybrid} approach utilizing recombination of two quarks and the subsequent fragmentation of the resulting diquark, which could provide enhanced baryon production while conserving jet structures, is suggested. The occurrence of different production mechanisms at intermediate $p_T$ might be studied further with fluctuations of the multiplicity associated with trigger particles in a narrow angular cone. To avoid a complete masking of these fluctuations by the unavoidable statistical fluctuations of the background,  constrained multiplicity distributions $P_J(I)$ can be used, which describe the probability that out of a fixed number of particles $J$ observed in a $180 ^{\circ}$ interval $I$ particles will end up in the narrower cone of $60 ^{\circ}$. A simple mathematical model of the different contributions to this multiplicity distribution is introduced. In this model, which ignores correlations in the background multiplicity, one finds that the shape of the multiplicity distribution varies significantly, if the relative fraction of events changes, where there are no correlated associated particles, i.e. no jet structures. Further investigations will be performed to understand the importance of background correlations.

\end{document}